\documentclass{PoS}
\usepackage{graphicx}
\usepackage{psfig}
\usepackage{epsfig}

\PoS{PoS(LAT2005)009}

\newcommand{\ve}{\varepsilon}
\newcommand{\beq}{\begin{equation}}
\newcommand{\eeq}{\end{equation}}
\newcommand{\beqa}{\begin{eqnarray}}
\newcommand{\eeqa}{\end{eqnarray}}
\newcommand{\ltil}{\tilde{\ell}}
\newcommand{\rtil}{\tilde{r}}

\newcommand{\Ltil}{\tilde{L}}
\newcommand{\ovr}{\over}
\newcommand{\barr}[1]{\not\mathrel #1}

\title{Quark mass dependence of baryon properties}

\ShortTitle{Quark mass dependence of baryon properties}

\author{\speaker{Ulf-G. Mei{\ss}ner}\thanks{This
research is part of the EU Integrated Infrastructure Initiative Hadron
Physics Project under contract number RII3-CT-2004-506078.
Work supported in part by DFG (SFB/TR 16 ``Subnuclear Structure of Matter'').}\\

        Helmholtz-Insitut f\"ur Strahlen- und Kernphysik\\
        Universit\"at Bonn\\
        Nu{\ss}allee 14-16\\
        D-53115 Bonn, Germany\\
        {\it and}\\
        Institut f\"ur Kernphysik (Theorie)\\
        Forschungszentrum J\"ulich\\
        D-52425 J\"ulich, Germany\\

        E-mail: \email{meissner@itkp.uni-bonn.de}}

\abstract{I discuss the quark mass dependence of various
          baryon properties derived from chiral perturbation theory.
          Such representations can eventually be used as chiral
          extrapolation functions when lattice data at sufficiently
          small quark masses become available. The quark mass dependence is
          encoded in loop and contact term contributions, the latter
          given in terms of low-energy constants.
          I stress the importance of utilizing phenomenological input
          to constrain a certain class of low-energy constants and discuss
          the ensuing theoretical uncertainty for various baryon observables,
          like the nucleon and the baryon octet masses, the nucleon isovector
          anomalous magnetic moment and the axial-vector coupling of the
          nucleon. I stress the role of resonance decoupling and present
          first results for the delta mass based on an effective field theory
          in which the nucleon-delta mass splitting is counted as a small
          parameter. I also discuss briefly the pion mass dependence of the
          nuclear force as derived from chiral nuclear effective field theory.
 }

\FullConference{XXIIIrd International Symposium on Lattice Field Theory\\

                 25-30 July 2005\\

                 Trinity College, Dublin, Ireland}

\begin{document}

\section{Approaching the chiral limit in QCD}
\label{sec:cl}
Chiral perturbation theory (CHPT) is the appropriate tool to analyze the quark mass
dependence of baryon properties. CHPT is the effective field theory (EFT) of
QCD, based on its symmetries and their 
realization~\cite{Weinberg:1978kz,Gasser:1983yg,Gasser:1984gg}. 
A central role in QCD is
played by the spontaneously and explicitly broken chiral symmetry for the
three light quark flavors up, down and strange. The vacuum is not invariant
under this global SU(3)$_L \times$SU(3)$_R$ symmetry and the spectrum
therefore contains
eight almost massless pseudoscalar Goldstone modes. These are not exactly
massless because the quark masses are non-zero but have very small
values compared to typical strong interaction scales.  The consequences of
this explicit and spontaneous chiral symmetry breaking can be analyzed in the
framework of CHPT.  S-matrix elements and transition currents are expanded
in small parameters - external momenta and quark (meson) masses with respect
to the chiral symmetry breaking scale $\Lambda_\chi \simeq 1\,$GeV.
Matter fields (that is all particles that are not
Goldstone bosons) can also be considered in CHPT. A special role is played
by the ground state baryons - they can not decay strongly and thus one is able
to consistently deal with the new mass scale related to these fields
-- the chiral limit value of the baryon mass is neither non-vanishing nor 
small compared to the scale of
chiral symmetry breaking $\Lambda_\chi$.  Resonances, may they be meson or
baryons, can not be incorporated so directly, one either has to resort to
resummation schemes or consider additional small parameters that do not vanish
for vanishing quark masses in the construction of the effective field theory. 
For reviews on CHPT including also matter fields I refer 
to~\cite{Meissner:1993ah,Ecker:1994gg,Bernard:1995dp,Pich:1995bw,Meissner:2000gh} 
and the foundations of CHPT are discussed in depth in~\cite{Leutwyler:1993iq}.

We now consider the approach to the chiral limit. For the following
discussion it is important to differentiate between the SU(2) and the SU(3)
chiral limit (c.l.) of QCD. More precisely, in the {\sl SU(2) c.l. we have 
$m_u, m_d \to 0\,$ with $m_s$ fixed at its physical value and the heavy quarks
decoupled}; while in the {\sl SU(3) c.l. $m_u, m_d, m_s \to 0\,$ with the heavy
quarks decoupled}. There are a few general observations about the chiral
limit of QCD. First, S-matrix elements exist in the chiral limit for arbitrary
momenta. Second, as it is well-established, the approach to the chiral limit
is non-analytic in the quark masses $m_q$, leading to the famous chiral 
logarithms and other non-analytic behavior of certain observables.  
Third, in the fundamental paper~\cite{Gasser:1979hf} a decoupling
theorem was derived. {\em It states that the leading chiral non-analytic terms stem 
from pion (Goldstone boson) one-loop graphs coupled to pions (Goldstone
bosons) or nucleons (ground state baryons).} This theorem has various immediate
consequences:  resonances like the $\rho$, the $\Delta$, $\ldots$ decouple to
leading order. This puts  severe constraints on the construction of EFTs with 
resonance fields. Related to this is the observation that chiral limit of QCD 
and the large $N_c$ limit do not commute -- we will come back to these issues later.

\section{One-loop representation of baryon observables}

We now consider a complete one-loop (fourth order) SU(2) calculation. Any
nucleon observable ${\mathcal O}$ -- given in terms of tree and loop diagrams --
takes the form
\beq
{\mathcal O} (M_\pi , F_\pi ; m_0 , g_A; c_i , d_i, e_i )~,
\eeq
where some of the quantities entering this expression are themselves quark
mass dependent. Specifically, the pion mass and the pion decay constant are
given by   $M_\pi^2 = (m_u + m_d) \, B_0 + {\mathcal O}(m_q^2)$ and
$F_\pi  = F_0 \, \left[ 1 + {\mathcal O}(m_q)\right]$, where $B_0$ is related
to the quark condensate and $F_0$ is the pion decay constant in the SU(2)
chiral limit. Further, $m_0$ is the nucleon mass in the SU(2) c.l. and $g_A$
is the leading order (dimension one) axial-vector pion-nucleon coupling, 
$g_A  = g_0 \, \left[ 1 + {\mathcal O}(m_q)\right]$. Also, the
$c_i, d_i, e_i$  are low-energy constants (LECs) of the dimension two,
three, and four chiral effective pion-nucleon Lagrangian ${\mathcal L}_{\rm eff}$,
\beq
 {\mathcal L}_{\rm eff} = {\mathcal L}_{\pi N}^{(1)} +  {\mathcal L}_{\pi N}^{(2)} + 
 {\mathcal L}_{\pi N}^{(3)} +  {\mathcal L}_{\pi N}^{(4)} + \ldots~, 
\eeq
where the superscript $(n)$ denotes the chiral dimension (the purely
pionic part of the Lagrangian is not made explicit here). The complete
effective Lagrangian  is given in~\cite{Fettes:2000gb}. For later use,
I display the dimension one and two terms:
\beqa\label{Leff}
\mathcal{L}_{\pi N}^{(1)} &=&\bar{\Psi}\biggl(  i\barr{D}-m_0 +\frac{1}
{2}{g_0}\barr{u}\gamma_{5}\biggr)  \Psi ~, \nonumber\\
\mathcal{L}_{\pi N}^{(2)}
  &=& \sum_{i=1}^7 \bar\Psi \, c_i\, O_i^{(2)}\, \Psi =
\bar\Psi \biggl( {c_1} \langle \chi_+ \rangle  
+ {c_2} \bigl(-\frac{1}{8m_0^{2}
}\left\langle u_{\mu}u_{\nu}\right\rangle D^{\mu\nu}+{\rm h.c.}\bigr)
+ {c_3} \,\frac{1}{2}\left\langle u\cdot u\right\rangle\nonumber\\
&& \quad + \, {c_4} \,\frac{i}{4}
\left[  u_{\mu},u_{\nu}\right]  \sigma^{\mu\nu}
 + {c_5} \, \widetilde{\chi}_{+} 
+ c_6 \,\frac{1}{8m_0}F_{\mu\nu}^{+}\sigma^{\mu\nu}
+ c_7 \, \frac{1}{8m_0}\left\langle F_{\mu\nu}^{+}\right\rangle \sigma^{\mu\nu}
\,\biggr) \Psi~,\nonumber\\
\mathcal{L}_{\pi N}^{(3)}
  &=& \sum_{i=1}^{23} \bar\Psi \, d_i\, O_i^{(3)}\,\Psi ~, \qquad
\mathcal{L}_{\pi N}^{(3)}
  = \sum_{i=1}^{118} \bar\Psi \, e_i\, O_i^{(4)}\,\Psi ~, 
\eeqa
with $U(x) = u^2(x)$ collects the pion fields 
$\pi (x) = \vec{\tau}^a \cdot \vec{\pi}^a(x)$, $\Psi (x)$ the isodoublet
nucleon field,  $u_\mu = i u^\dagger
D_\mu U u^\dagger =  - i\partial_\mu \pi /F_0 + \ldots$ and $D_\mu$ the chiral
covariant derivative. The $O_i^{(n)}$ are monomials of chiral dimension
$n$. Furthermore, $\chi_+$ parameterizes the
external scalar source including the quark masses, $\chi_+ \sim 2B_0 {\rm
  diag}(m_u,m_d) + \ldots$, $F_{\mu\nu}^{+}$ is the chiral and gauge covariant
photon field strength tensor, and $\langle ~\rangle$ denotes the trace in
flavor space (for more details, see~\cite{Bernard:1995dp,Fettes:2000gb}).
{\em The LECs can be grouped into two distinct classes.} Class~I are the so-called 
{\bf dynamical} LECs $\sim \partial_\mu^2 , \partial_\mu^4, \ldots$ which are
non-vanishing in the chiral limit. Many of these are accessible via
phenomenology, that is through the analysis of scattering data or decay processes.
Class~II LECs, the so-called {\bf symmetry breakers} $\sim m_q ,  m_q^2,
 m_q \partial_\mu^2 , \ldots$, vanish in the c.l. and they can best be
 obtained by varying the quark masses, i.e. in lattice simulations (see also
the discussion in~\cite{Gasser:2003cg}). Note that the mixed LECs
parameterizing operators with quark mass insertions and derivatives are
counted as symmetry breakers because at fixed pion mass these can be absorbed
in the values of certain dynamical LECs (with the exception of the leading
mixed LECs at dimension three). The dimension two effective
Lagrangian in Eq.~(\ref{Leff}) has two symmetry breakers and five dynamical
LECs, from the latter two can only be determined in the presence of external fields.
At dimension three, we have 4 symmetry breakers and 19 dynamical LECs - in
fact, the symmetry breakers are necessarily of the mixed type and thus are
easier available from phenomenological analysis as most other LECs of that
type. Finally, at dimension four, we have 25 symmetry breakers (most of them
of mixed type) and 93 dynamical LECs (most of them containing external
fields). For more details, see \cite{Fettes:2000gb}. As we will see in the
following, while the total number of LECs increases rapidly, this is not the
case for most observables -- this will become evident when I discuss the
nucleon mass, the isovector magnetic moment or the axial-vector coupling constant.

\section{Analysis of baryon properties}

Chiral perturbation theory supplies the quark mass expansion of any given
observable to the required accuracy - provided the expansion parameter
is sufficiently small. There is an important interplay between chiral
loops and the LECs parameterizing the contact interactions. Mostly, one
employs dimensional regularization to perform the necessary renormalization of
the loop contributions in a symmetry-preserving manner
(although other schemes are viable, too). While
an observable ${\mathcal O}$ does not depend on the scale of dimensional 
regularization $\lambda$, that is $d{\mathcal O}/d\lambda = 0$, this is in general
not the case for the individual contributions from the loops or the counter
terms. An example of this is the isovector charge radius of the nucleon
as discussed in~\cite{Bernard:2003rp}. It is therefore absolutely necessary
to include {\em all} terms in harmony with the underlying symmetries, and
not just the chiral logs or other non-analytic terms. Also, it has become
evident over the years that in baryon CHPT one has to perform complete
one-loop (fourth order) calculations to obtain a precise enough representation
at physical pion/kaon masses (in SU(3), this statement is not true in
general). For reviews, see~\cite{Meissner:2003sb,Borasoy:2004ty}.

Before continuing, let me make the following \underline{{\sl disclaimer}}: 
I will discuss 
continuum extrapolations, but no finite lattice spacing or finite volume effects
Also, I will not discuss (partial) quenching, that means all lattice data are 
taken as full QCD. Further, I will be mostly interested in the range of 
applicability (theoretical uncertainty) rather than precision fits -- for reasons 
that become apparent in the following -- and finally, I eschew models here -- 
\underline{{\sl end of disclaimer}}.

\subsection{Quark mass expansion of the nucleon mass in SU(2) and SU(3)}
\label{sec:mN}
Consider the expansion of the nucleon mass in the light quark mass
$\hat m = m_u = m_d$ with  $m_s$ fixed. By means of two-flavor CHPT, this is
mapped onto an expansion in the pion mass. Note that we consider 
strong isospin violation $(m_u \neq m_d)$ later. The fourth order calculation
gives~\cite{Steininger:1998ya}
\beq\label{mN}
m_N = m_0  - 4{c_1} M_\pi^2- {\frac{3g_A^2 M_\pi^3}{32\pi F_\pi^2}} +
 {k_1} \, M_\pi^4 \ln {M_\pi \over m_N} + {k_2}\, M_\pi^4 +
{\cal O}(M_\pi^5)~,
\eeq
with $m_0$  the nucleon mass in the chiral SU(2) limit and $k_1$ and $k_2$
are combinations of second $(c_1, c_2, c_3$)  and fourth order LECs ($e_1$),
$k_1 = -({3}/{32\pi^2F_\pi^2}) ( -8{c_1} + {c_2} + 4{c_3} +
{g_A^2}/{m_N})$ and ${k_2} = -4{e_1}  + ({3}/{128\pi^2F_\pi^2}) (
{c_2} -  {2g_A^2}/{m_N})$, respectively. The scale of dimensional
regularization was set equal to the nucleon mass, $\lambda = m_N$. 
For simplicity, I have neglected in Eq.(\ref{mN}) the difference between
the physical value of the pion mass and its leading term in the quark mass
expansion, $M_\pi^2 = M^2 [1 +{\cal O}(m_q)]$ (for details, see 
e.g.~\cite{Becher:1999he}). Note also that the dimension
two LECs are scale-independent in any mass-independent regularization scheme,
while the fourth order LEC is scale-dependent,  $e_1 = e_1 (\lambda)$.
It is important to realize that the 
LECs $c_1, c_2, c_3$, and $e_1$ are strongly constrained from pion-nucleon 
scattering and to some extent form peripheral nucleon--nucleon scattering.
More specifically, the chiral expansion of the $\pi N$ amplitude converges
best inside the Mandelstam triangle, however, to determine the values
of the LECs, one must compare to a dispersive analysis of the scattering
data. This was done in~\cite{Buettiker:1999ap}. Alternatively, one can use the
chiral representation of the amplitude in the physical region close to
threshold, here one is faced with certain inconsistencies in the data basis.
The most detailed fits based on these data were performed
in~ \cite{Fettes:1998ud}. One can also pick out
certain threshold parameters that are particularly sensitive to  certain
LECs, as it was done in~\cite{Bernard:1996gq}. In addition, the peripheral
partial waves in elastic nucleon-nucleon are sensitive to the two-pion
exchange and thus to some of the $c_i$. In this case, one has a large data basis
but must subtract the dominant one-pion exchange, 
see~\cite{Rentmeester:2003mf,Epelbaum:2003gr} (see also
Ref.~\cite{Entem:2003ft} -- which, however, does not enter my average). Putting all
this information together, we obtain for the $c_i$ $(i=1,\ldots,4)$ 
(all values in GeV$^{-1}$)
\beq\label{cival}
c_1 = -0.9^{+0.5}_{-0.2}~,~~ c_2 = 3.3 \pm 0.2~,~~ c_3 = -4.7^{+1.2}_{-1.0}~,~~
c_4 = 3.5^{+0.5}_{-0.2}~,~~ -2c_1+ \frac{c_2}{4} + c_3 = -2.1^{+1.3}_{-1.1}~.
\eeq
A few remarks are in order. First, the symmetry breaker $c_1$  and the
isoscalar combination (last number) are least accurately determined -- as
expected. Also, the LEC $e_1$ is only badly determined from $\pi N\to \pi N$
in the threshold region, the analysis of~\cite{Fettes:2000xg} gives a negative 
value of natural size. It is also important to notice that the numerical
values of the LECs $c_i$ can be well understood in terms of resonance
saturation (baryon and meson resonance excitations), in particular, the
fairly large values of $c_{2,3}$ ($c_4$) are generated mostly by 
$\Delta (1232)$ ($\rho$) exchange~\cite{Bernard:1996gq}. Before discussing
the quark mass dependence of the nucleon mass as given by Eq.~(\ref{mN}), it
is important to point out that for the physical pion mass, the higher order
corrections are small. More precisely, for the central values of the LECs
given in  Eq.~(\ref{cival}) (and using $e_1 = -1\,$GeV$^{-3}$), the contribution
quadratic, cubic and quartic in $M_\pi$ amounts to $70$, $-17$ and $-4\,$MeV,
respectively. It was even shown in \cite{McGovern:1998tm} 
that the fifth order corrections are tiny (at the physical point). 
I will return to this issue of convergence later.

\begin{figure}[htb]
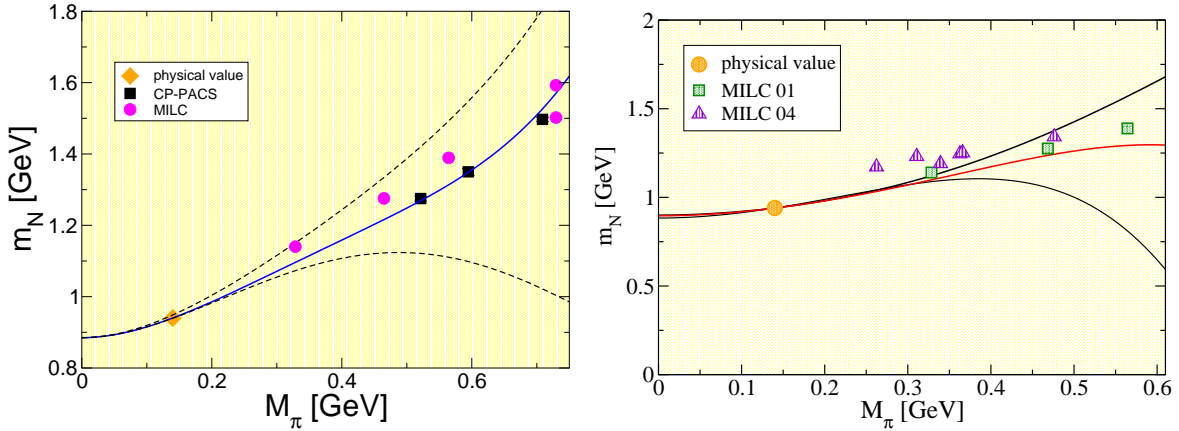

\includegraphics*[width=7.5cm]{msu2new.eps}~~~
\includegraphics*[width=7.7cm]{msu3new.eps}
\caption{Pion mass dependence of the nucleon mass. Left panel: SU(2) analysis.
The blue solid line refers to the best fit and the dashed black lines to the
theoretical uncertainty as discussed in the text. Red circles: 
MILC data~\protect\cite{Bernard:2001av}, black squares: CP-PACS data~\cite{AliKhan:2001tx}.
Right panel: SU(3) analysis.
The red line gives the best fit as discussed in the text and the black lines
give the theoretical uncertainty under the constraint that the nucleon mass
takes its physical value for the physical quark masses.
Green squares: MILC 2001 data~\protect\cite{Bernard:2001av}, purple triangles: MILC
2004 data ~\protect\cite{Aubin:2004wf}
}
\label{fig:mN}
\end{figure}
\noindent 

In the left panel of Fig.~\ref{fig:mN} we show the results from 
Ref.~\cite{Bernard:2003rp}. A best fit to the
MILC 2001~\cite{Bernard:2001av} and the CP-PACS data~\cite{AliKhan:2001tx}  
is obtained with $c_1 = -0.9, c_2 = 3.2, c_3=-3.5$ (all in
GeV$^{-1}$) and $ e_1 (m_N) = -1\, {\rm GeV}^{-3}$.  Note that this is not a chiral
extrapolation in the true sense of the word because the constraint that 
the physical value of $m_N$ is obtained  at the physical pion mass was
imposed. Note also that the MILC data are from an  SU(3) simulation, however, 
they have also given an SU(2) reference value for one pion mass that differs only little
from the corresponding SU(3) result.  
The resulting SU(2) c.l. value of the nucleon mass is $m_0 \simeq 0.88\,$GeV.
The dashed lines in the figure reflect the theoretical uncertainty from the
variation of the LECs, consequently, we obtain a  moderate/large theoretical 
uncertainty for $M_\pi$ above 400/550~MeV. Clearly, at higher pion masses the
higher order terms become too important for such fits to make sense. For
related work on the nucleon mass in SU(2),
see~\cite{Procura:2003ig,AliKhan:2003cu}. The theoretical uncertainty in the
determination of the pion mass dependence of the nucleon mass is also
discussed in~\cite{Beane:2004ks}.

Next, I consider the extension to the three-flavor case. Before discussing the
nucleon mass, some general remarks are in order. As compared to the SU(2) analysis,
one has more fields and operator structures in the EFT, and consequently, more LECs.
As examples, consider the leading order axial meson-baryon coupling in SU(2) and SU(3)
\beq
\textstyle\frac{1}{2} g_A \, \bar\psi u_\mu \gamma^\mu \gamma^5 \psi \to
{D} \, \langle \bar{B}  \gamma^\mu \gamma^5\{ u_\mu,B\}\rangle +
{F} \, \langle \bar{B}  \gamma^\mu \gamma^5 [ u_\mu,B ]\rangle~,
\eeq
or the leading symmetry breakers
\beq
c_1  \, \bar\psi \langle \chi_+ \rangle \psi \to
{b_0} \, \langle \bar{B} B\rangle +
{b_D} \, \langle \bar{B} \{ \chi_+ ,B\}\rangle +
{b_F} \, \langle \bar{B} [ \chi_+ , B ]\rangle~,
\eeq
with $\Psi$ the nucleon doublet and $B$ the baryon octet coupled to the pion
triplet and the Goldstone boson octet, respectively. Note that the LEC $b_0$
can be absorbed in the value of the octet baryon mass in the chiral limit, but
need to be kept separately when one also analyses the pion- and kaon-nucleon
sigma terms~\cite{Bernard:1993nj}. These various operators
in SU(2) and SU(3) are  related by {\em matching conditions} -- in the SU(2)
case the strange quark effects are buried in certain local operators. Such
matching relations have been worked out in~\cite{Frink:2004ic}, to leading
order, one finds, e.g. 
\beq
\begin{array}{ll}
{\rm nucleon~ mass~ in~ the~ SU(3)~ chiral~limit:} & \quad
\tilde m_0 = m_0 \left[ 1 +{\mathcal O} (m_s) \right]~, \\
{\rm leading~ axial~ coupling:} & \quad g_A = D + F + {\mathcal O} (m_s)~,\\
{\rm leading~ symmetry~ breaker:} & \quad 
c_1 =b_0 + \frac{1}{2} \left( b_D + b_F \right) + {\mathcal O} (\sqrt{m_s})~.
\end{array}
\eeq
The full matching conditions to fourth order in the chiral expansion are
given in ~\cite{Frink:2004ic}.
These are important constraints that should be implemented in any SU(3)
analysis. I return to the nucleon mass. Ref.~\cite{Frink:2004ic} contains
the complete fourth order expressions (including isospin breaking terms) for the
baryon octet masses and sigma terms. Based on that work,
in~\cite{Frink:2005ru} fits to the MILC 2001 data based on two different
regularization schemes (cut-off and dimensional regularization) were
performed. Again, the constraint to obtain the physical value of the nucleon
mass at the physical quark masses was imposed. The dimension two LECs called
$b_i$ were taken from the earlier work in~\cite{Borasoy:1996bx} and the
three combinations of the dimension four LECs (called $d_i$) were 
varied to obtain a best description of the three-flavor MILC 2001 data, 
see the red line in the right panel of
Fig.~\ref{fig:mN}.  In the figure are also shown the 
MILC 2004 data~\cite{Aubin:2004wf}
at lower quark masses  -- these can not
be fitted with the constraint (as already noted by the MILC collaboration 
using a simplified mass formula).  Taking into account
also the kaon mass dependence of the nucleon mass and the uncertainty
due to the MILC 2004 data, we obtain the following ranges for various
(isoscalar) quantities:
\beq
\begin{array}{lc}
{\rm SU(3)~ c.l.~ value~ of~ m_N:} & \quad 710~{\rm MeV} \lesssim
  \tilde{m}_0 \lesssim 1070~{\rm MeV}~, \\
{\rm Pion-nucleon~ sigma~ term:} & \quad 39.5~{\rm MeV} 
\lesssim \sigma_{\pi N}(0) \lesssim  46.7~{\rm MeV}~,\\
{\rm Strangeness~ fraction:} & \quad 0.07 \lesssim y \lesssim 0.22~,
\end{array}
\eeq
with $\sigma_{\pi N}(0) =  \langle N(p)|\hat{m} (\bar uu +\bar
dd)|N(p)\rangle$ (with $N(p)$ a nucleon state of momentum $p$)
and the strangeness fraction in the proton,  
$y = 2 \langle p| \bar s s|p\rangle /\langle p|\bar uu +\bar dd|p \rangle$,
can be obtained from $\sigma_{\pi N} (0) =
\sigma_0 /(1-y)$ and using $\sigma_0 = (37\pm 6)\,$MeV from \cite{Borasoy:1996bx}.
It was  concluded in~\cite{Frink:2005ru} that for pion masses less 
than 400~MeV the convergence of the chiral expansion is fine and that it is 
acceptable for masses below 550~MeV. This agrees
with the findings of the SU(2) analysis in~~\cite{Bernard:2003rp}.

\subsection{Quark mass expansion of the baryon octet masses}

\begin{figure}[tb]
\includegraphics*[width=5.5cm,angle=270]{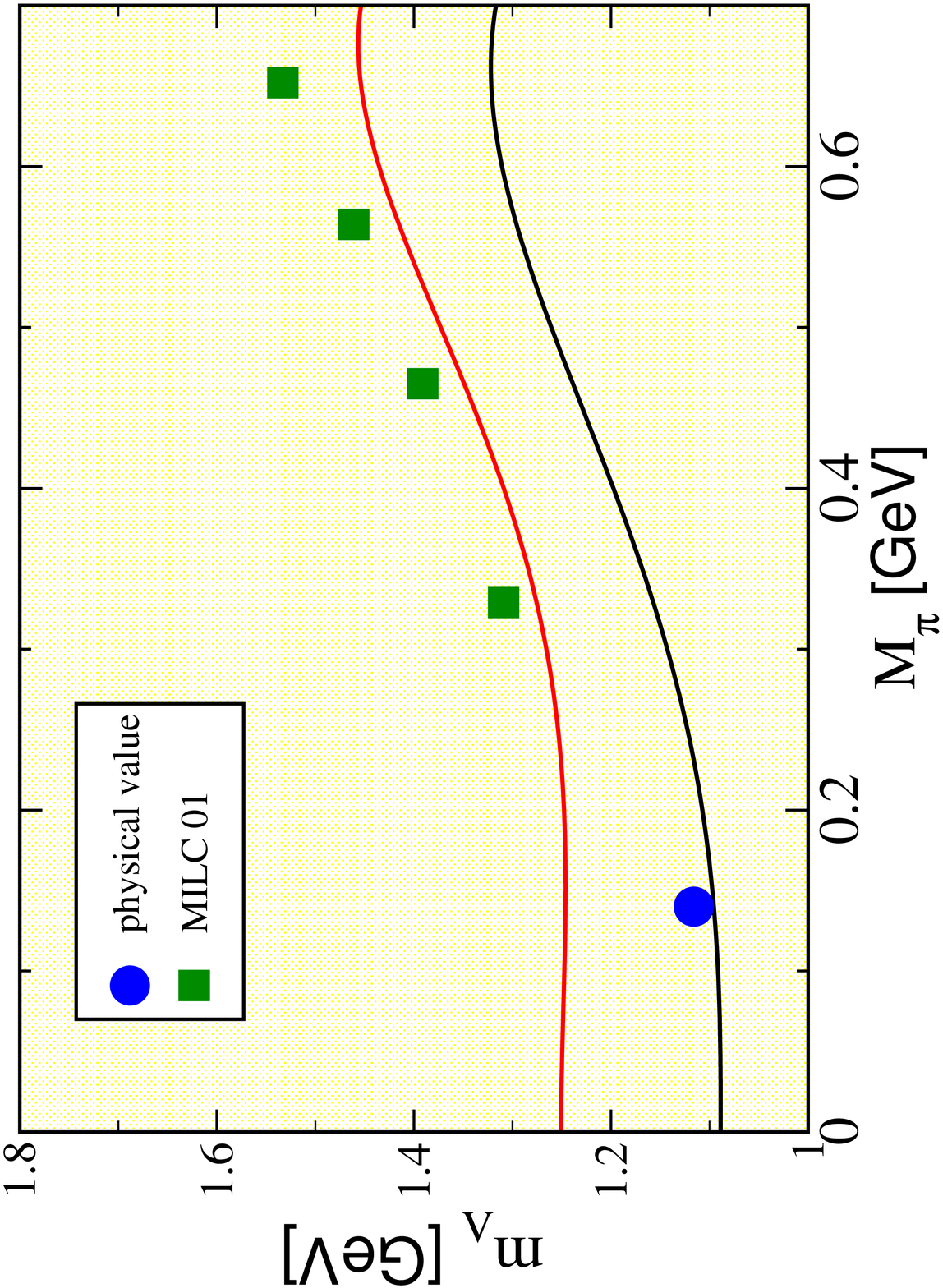}
\includegraphics*[width=5.5cm,angle=270]{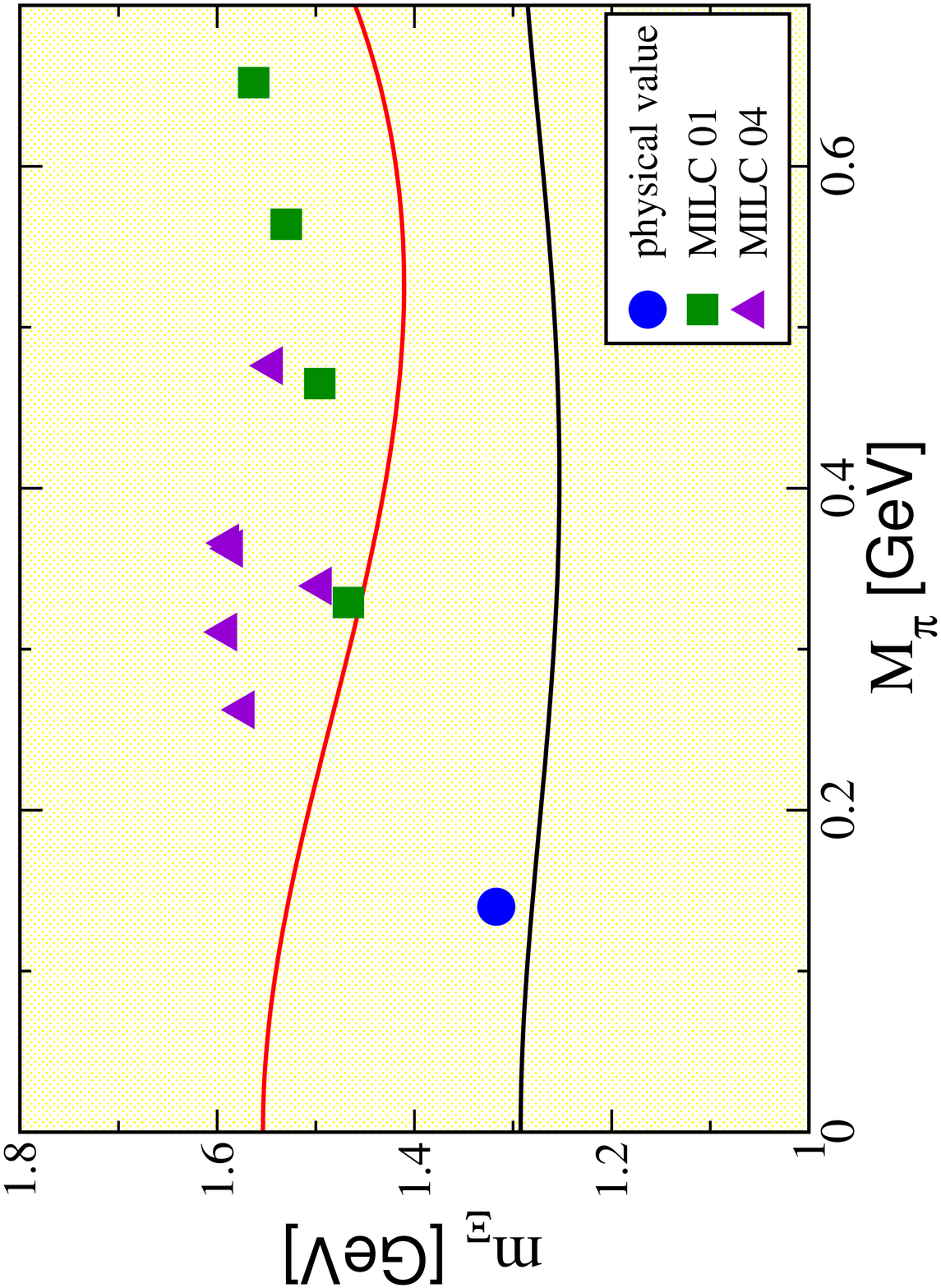}
\caption{Pion mass dependence of the $\Lambda$ mass (left panel)
and of the $\Xi$ mass (right panel). Green boxes/purple triangles:
MILC 2001/2004 data. The blue circle denotes the physical value
of the corresponding baryon mass. For further details, see the text.
}
\label{fig:hyp}
\end{figure}
\noindent
We now consider the other octet members with strangeness ($\Lambda,
\Sigma,\Xi$) based on the work of
Ref.~\cite{Frink:2005ru}. In that paper, the fourth order LECs
were fitted to the nucleon mass only, that means the Gell-Mann--Okubo 
relation for the baryon masses was not imposed. Consequently,
some of the masses come out somewhat off their empirical values. More
precisely, while the $\Sigma$ mass is well
reproduced, the $\Lambda$ and $\Xi$ masses come out by about $10 - 15$\% too
high. To get a  handle on the theoretical accuracy, we also use the values
for the $d_i$ from \cite{Borasoy:1996bx}, in that case all masses are exactly reproduced.
The resulting prediction for the pion mass dependence of the $\Lambda$ and
the $\Xi$ in comparison to the MILC data are shown in Fig.~\ref{fig:hyp}.
The red/black lines refer to the optimal set of the LECs from the 
nucleon mass fit/to the LECs from \cite{Borasoy:1996bx}. We note in particular 
that the pion mass dependence for the $\Xi$ is much flatter as one 
would expect from the MILC data. This is not unexpected -- the $\Xi$ 
only contains one valence light quark and should thus be less sensitive to 
variations in the pion mass. Clearly, one
could improve this description by fitting directly to these particles.
Still, the pion mass dependence of the $\Xi$ as given by the MILC data
is a mystery to be resolved. 

So far, I have considered the isospin limit $m_u = m_d$. I briefly discuss the
implications of unequal light quark masses for the strong neutron--proton 
mass difference. In the  SU(2) case, it is given to leading order by
\beq\label{mnp}
(m_n -m_p)^{\rm strong} = 4c_5 B_0 (m_u-m_d) + {\mathcal O}((m_u-m_d)^2)~
\eeq
with $c_5$ the  dimension two LEC accompanying the  operator $\tilde{\chi}_+ = \chi_+ -
\frac{1}{2}\langle\chi_+\rangle \sim B_0 (m_u - m_d)$, see also
Eq.~(\ref{Leff}). Utilizing the 
 Cottingham sum rule to estimate the electromagnetic splitting, one
finds~\cite{Bernard:1996gq}
\beq  
c_5 = -0.09\pm 0.01\,{\rm GeV}^{-1}~. 
\eeq
The small value of this LEC is a reflection of the suppression of isospin
violation in QCD, which scales as $(m_u -m_d)/\Lambda_\chi$.
(Note that Weinberg obtained a slightly larger value in his seminal paper because
he employed SU(3) relations~\cite{Weinberg:1977hb}). The fourth order corrections 
to Eq.~(\ref{mnp}) have been worked out, see e.g.~\cite{Muller:1999ww}.
In SU(3), the expression for the strong neutron--proton mass 
is much more complicated, since one has to treat 
$\pi^0-\eta$ and $\Lambda-\Sigma^0$ mixing in terms of the parameter
$\tan 2\epsilon = ({\sqrt{3}}/{2})({m_u-m_d})/({m_s-\hat m})$. Again,
one can derive a matching condition~\cite{Frink:2004ic}, to leading order
it reads (for related work, see the pioneering analysis in~\cite{Gasser:1980sb} 
and also~\cite{Tiburzi:2005na})
\beq 
c_5 = b_D + b_F + {\mathcal O}(m_s, m_s \ln m_s)~.
\eeq
Needless to say that for obtaining  precise mass splittings, one 
needs to consider also the electromagnetic corrections
(for an early analysis in the meson sector, see~\cite{Duncan:1996xy}).

\subsection{Quark mass expansion of other nucleon properties}
\label{Nprops}

Besides the nucleon mass and sigma term, chiral extrapolations for
various other nucleon observables based on chiral perturbation theory 
(or extensions thereof) have been  considered in the literature, 
e.g. nucleon magnetic moments and the  
electromagnetic radii~\cite{Hemmert:2002uh,Ashley:2003sn,Gockeler:2003ay} 
or the isovector axial-coupling 
constant~\cite{Hemmert:2003cb,Beane:2004rf,Fuhrer}. I will not review these
in detail but rather make some more general remarks on the present status
of  extrapolation functions derived from baryon chiral perturbation
theory. There are two important issues which require special attention:

\vspace{-0.2cm}

\begin{itemize}
\item[\fbox{1}]~Given the scarcity of data at low pion
masses (say below $M_\pi \simeq 400$~MeV), one should perform global fits 
to a variety of observables at sufficiently small quark masses. The
important point is that the  LECs -- the parameters of these fits -- 
relate many observables, they are the same for all processes and 
can therefore \underline{not} be determined  independently.
Furthermore, one should incorporate as much  phenomenological input
as possible, in particular for the dynamical LECs, see the discussion of
the nucleon mass in the preceeding sections. 
Needless to say that one must be in a regime where higher order terms 
stay sufficiently small. Attempts to go beyond this regime are necessarily
model-dependent, see e.g.~\cite{Leinweber:2005xz} and references therein.

\vspace{-0.2cm}
%
%
\item[\fbox{2}]~Results should be independent of the regularization scheme,
since these can differ only by higher order terms if symmetries are properly
implemented. For example, the so-called infrared regularization (IR) 
of baryon CHPT~\cite{Becher:1999he}
resums all kinetic energy insertions $\sim {\vec p\,}^2/2m_N$ whereas the 
latter are expanded  order-by-order in the heavy baryon approach (there are
a few circumstances for which the heavy baryon approach does not converge, but
these are not of relevance here). Similarly, cut-off schemes or
finite-range-regulators as employed by the Adelaide group, 
see e.g.~\cite{Leinweber:2003dg}, resum certain higher order corrections (if implemented
properly). For small pion masses as defined above, all these schemes
must give the same result.
\end{itemize}

\vspace{-0.2cm}
\begin{figure}[tb]
~~~~~~~\includegraphics*[width=4.9cm]{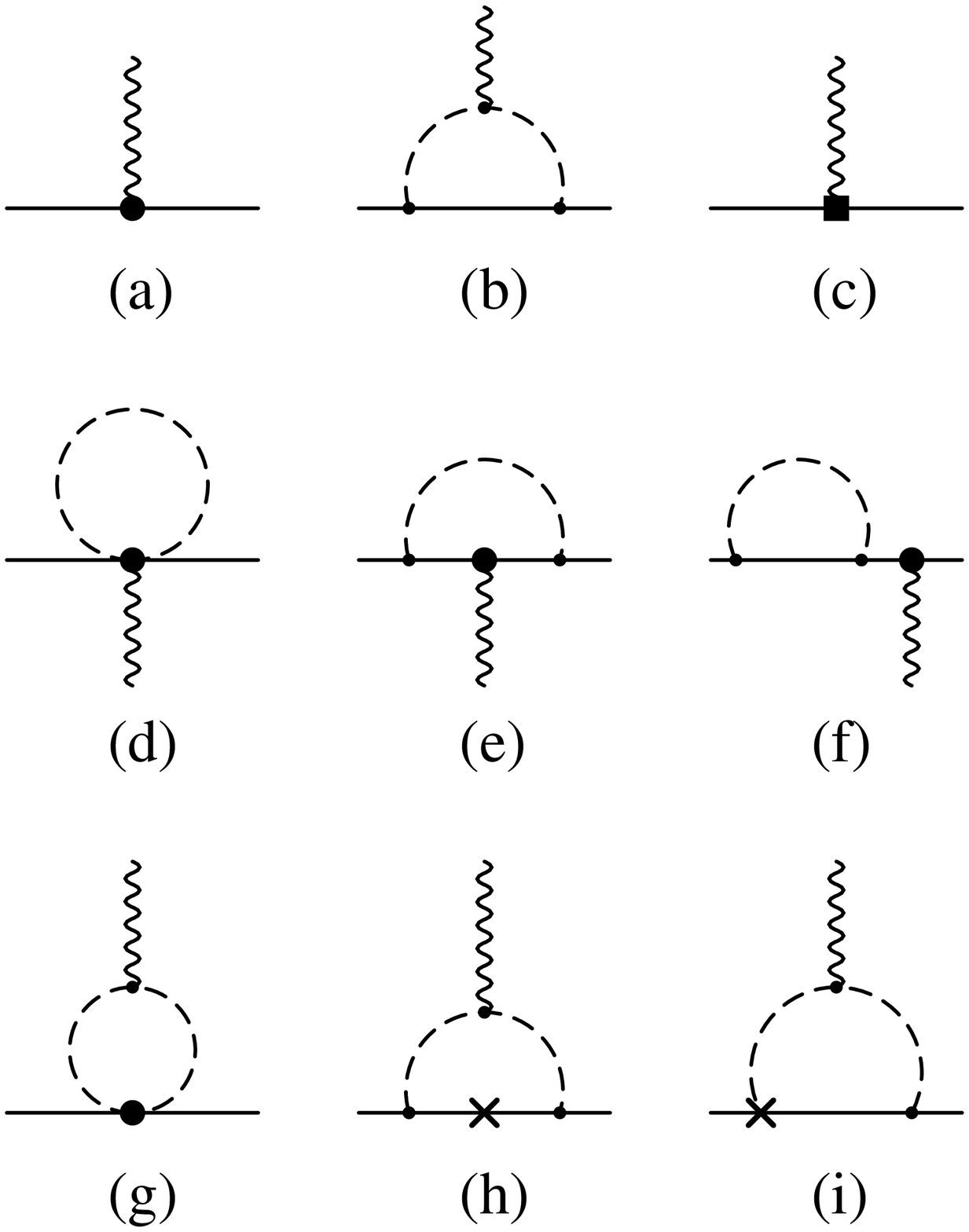}~~~~~~~~~~
\includegraphics*[width=8.1cm]{kappadep.eps}
\caption{Left panel: Second (a), third (b) and fourth (c-i)
order diagrams contributing
to $\kappa_v$. The solid/dashed/wiggly lines stand for
nucleons/pions/photons, respectively. The $\bullet$
(${\blacksquare}$) denotes a dimension two (four) insertion.
The $\times$ denotes an dimension two insertion with fixed coefficients.
Right panel: Pion mass dependence of $\kappa_v$
as described in the text: The upper (lower)
line corresponds to the c.l. (physical) values
of the pion decay constant and the nucleon mass.
The diamond represents the physical value of
$\kappa_v$. The data are from QCDSF~\protect\cite{Gockeler:2003ay}.}
\label{fig:kappav}
\end{figure}
\noindent
Let me  illustrate these issues on the example of the nucleon isovector anomalous
magnetic moment $\kappa_v$. It has been worked out to fourth order 
in~\cite{Kubis:2000zd} (the pertinent Feynman diagrams are shown in the
left panel of  Fig.~\ref{fig:kappav})
\beq\label{kappav}
\kappa_v = c_6 -16 m_N M_\pi^2 e_{106} + \kappa_v^{\rm loop,3} + \kappa_v^{\rm loop,4}~,
\eeq
where I have made explicit the contributions from the dimension two $\sim c_6$
(diagram (a) in  Fig.~\ref{fig:kappav}) and four $\sim e_{106}$ contact 
interactions (diagram (c) in  Fig.~\ref{fig:kappav}). Here, $c_6$ is nothing but the
anomalous magnetic moment in the chiral limit, $\kappa_v^0 = c_6$. The fourth
order loop contribution depends explicitely on the dimension two LECs 
$c_2$ (the tadpole graph (g) in  Fig.~\ref{fig:kappav}),
$c_4$ (the tadpole graph (d) in  Fig.~\ref{fig:kappav}),
and $c_6$ (diagrams (e) and (f) in Fig.~\ref{fig:kappav}). 
Two of these are constrained from pion-nucleon scattering, cf. Eq.~(\ref{cival}). 
The expression for $\kappa_v$ contains the nucleon mass $m_N$ and the pion
decay constant $F_\pi$ (in the loop contributions). These quantities depend
themselves on the pion mass, $m_N = m_N (M_\pi)$ and $F_\pi = F_\pi (M_\pi)$.
To the order we are working, one can always replace the parameters in the
effective Lagrangian ($m_0$ and $F_0$, the pertinent c.l. values, see e.g.
Eq.~(\ref{Leff})) by their physical values, the
differences being of higher order in the chiral expansion. Thus, 
replacing $\{ m_0, F_0\}$ by $\{ m_N, F_\pi\}$ in the formula 
Eq.~(\ref{kappav}) is a way of investigating the convergence of the series,
and I will do this here for $\kappa_v$.\footnote{I am grateful to V\'eronique
  Bernard for providing me with these results.} The explicit expression 
for the pion mass expansion of $m_N$ is given in Eq.~(\ref{mN}) and 
$F_\pi$ to ${\mathcal O}(M_\pi^4)$ can be taken from~\cite{Colangelo:2003hf}
(see also Ref.~\cite{Bijnens:1997vq} and references therein)
\beq
{F_\pi\ovr F_0} = 
1+X\Big[\Ltil+\ltil_4\Big]+X^2\Big[-{3\Ltil^2 \ovr4}
+ \Big(-{7\Ltil\ovr6}\ltil_1-{4 \ltil_2 \ovr3} + \ltil_4-{29\ovr12}\Big)
+{\ltil_3\ltil_4\ovr2} - {\ltil_1 \ovr12} - {\ltil_2 \ovr3} - {13\ovr192}
+\rtil_F(\lambda)\Big]
\label{eq:Fpi}
\eeq
where $\Ltil=\log(\lambda^2/M_\pi^2)$, $X=M_\pi^2/(16\pi^2 F_0^2)$ and $\rtil_F (\mu)$
is a combination of dimension six LECs. I use $\lambda = 0.5\,$GeV and 
$\rtil_F(\lambda =0.5\,{\rm GeV}) = 3$ in the following. 
The NLO LECs $\ltil_i$ are tabulated in~\cite{Colangelo:2003hf}. 
Guided by the trend of the QCDSF 
data~\cite{Gockeler:2003ay} and imposing the constraint that $\kappa_v =
3.71\,$[n.m.] at the physical pion mass, we set $c_6 = 5$ and $e_{106} =
0.45$. The resulting theoretical uncertainty due to the two choices for the
pion decay constant and the nucleon mass are shown in Fig.~\ref{fig:kappav}.
As before, below pion masses of about 400~MeV, the theoretical uncertainty
is modest, but above it quickly increases -- so that a model-independent
extraction of $\kappa_v$ using the particular data shown in the figure is not possible.

Let me now consider  the axial-vector coupling $g_A$, which was
much discussed in the recent literature, see 
e.g.~\cite{Hemmert:2003cb,Beane:2004rf,Fuhrer}. I present here
some first results of an on-going study with V\'eronique Bernard~\cite{Bernard05}. 
There are two main issues concerning this particular observable. 
First, to my opinion, the existing data are at too high pion masses   
to draw definite conclusions about
the range of applicability of the chiral extrapolation functions or the
role of explicit spin-3/2 degrees of freedom.  Second, there is a much
more direct problem with the pion mass dependence of $g_A$.
It has been worked out to fourth order first in~\cite{Kambor:1998pi}.
Using the standard form of the effective pion-nucleon Lagrangian
(see Eq.(\ref{Leff}) and Ref.~\cite{Fettes:2000gb}) it reads
\beqa\label{gAeq}
g_A &=& g_0 \, \left[ 1  + \frac{4M_\pi^2}{g_0} \left( {d}_{16} (\lambda)
 - \frac{\frac{1}{2}g_0 + g_0^3}{16\pi^2F_0^2}\ln\frac{M_\pi}{\lambda}\right)
- \frac{g_0^2M_\pi^2}{16\pi^2F_0^2} \right. \nonumber\\
&& \left. \qquad\qquad\qquad + \frac{M_\pi^3}{24\pi F_0^2m_0}
\left(3+3g_0^2-4m_0c_3+8m_0c_4\right) \right] + {\mathcal O}(M_\pi^4)~.
\eeqa
The dimension three coupling constant ${d}_{16}$ can be determined
from the process $\pi N \to \pi\pi N$~\cite{Fettes:1999wp}, however, not
very accurately. For a typical value of the LEC $\bar d_{16} = -1.76\,$GeV$^{-2}$, 
the correction quadratic in the pion mass is $\Delta g_A^{(2)} \simeq 0.16$, 
which is of typical size for an SU(2) correction (note that $\bar{d}_{16}$ is
the renormalized value of $d_{16}$ at $\lambda = M_\pi$, see~\cite{Fettes:1998ud}).
It was, however,  already pointed out in~\cite{Kambor:1998pi} that the
corrections $\sim M_\pi^3$ are already quite sizeable for the physical 
pion mass. Using the central values of the LECs $c_3$ and $c_4$, one
finds $\Delta g_A^{(3)} \simeq 0.32$, using $F_0 =
86.5\,$MeV~\cite{Colangelo:2003hf}, $m_0 = 880\,$MeV~\cite{Bernard:2003rp}
and $g_0 =1.2$. This is a fairly large correction for an SU(2) quantity.
It can be traced back mostly to the particularly large combination of the
dimension two LECs $-4m_0c_3+8m_0c_4 \simeq 41$, and thus one can not expect
this representation to be very accurate. 
\begin{figure}[htb]
\centerline{\psfig{file=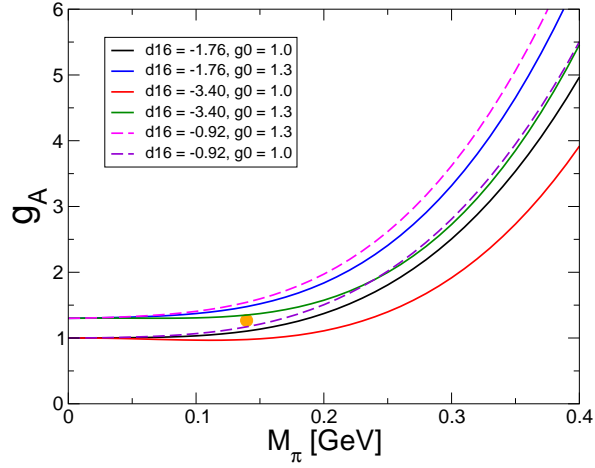,width=7.5cm,angle=270}}
\caption{Pion mass dependence of the axial-vector coupling for
various values of the LEC  $\bar{d}_{16} = -0.92, -1.72, -3.4\,$GeV$^{-2}$ 
and the chiral limit value of $g_A$, $g_0 = 1.0$ and $1.3$, respectively.
The filled orange circle denotes the physical value of $g_A$ at the physical
pion mass.
\label{figgA}
}
\end{figure}
\noindent This is further corroborated in
Fig.~\ref{figgA}, where the pion mass dependence of $g_A$ is shown for
various values of the LEC $\bar d_{16}$ and two values of 
 $g_0 = 1.0$ and $g_0 = 1.3$, respectively. The strong
increase of $g_A (M_\pi)$ with growing $M_\pi$ is due to the large correction
cubic in the pion mass -- this clearly narrows the window for applying 
Eq.(\ref{gAeq}) as a chiral extrapolation functions to even smaller pion
masses than it is the case for the nucleon mass or the isovector magnetic
moment discussed earlier. It should also be stressed that using an
extrapolation function based only on the third order calculation (i.e. including the
terms up to quadratic order in $M_\pi$) is insufficient - one does not even
capture the largest correction at the physical value of the pion mass. On the 
other hand, it is conceivable that even higher order corrections are again 
of natural size -- this, however, needs to be checked by an explicit calculation.
A more detailed discussion of these issues will
be given in~\cite{Bernard05} (see also~\cite{Hemmert:2003cb}).

\subsection{Quark mass expansion of the $\Delta$  mass}

The $\Delta (1232)$  is the most important baryon
resonance. It is almost degenerate in mass with the nucleon
and couples strongly to pions, nucleons and photons. It was therefore
argued early that spin-3/2 (decuplet) states should be included in baryon
chiral perturbation theory \cite{Jenkins:1991es}. In that paper  and 
subsequent works use was made of the heavy baryon approach, which treats 
the baryons as static sources. However, special care has to be taken about the
decoupling of resonances in the chiral limit, 
as discussed in Sec.~\ref{sec:cl} - this is frequently ignored in literature. 
This  approach can be systematized by counting the nucleon-delta mass 
splitting as an additional small parameter, so 
one expands in the generic small parameter $\ve$~\cite{Hemmert:1997ye}, with
\beq
\varepsilon \in \Bigg\{ {p\ovr\Lambda_\chi}, {M_\pi\ovr\Lambda_\chi}, 
{m_\Delta - m_N \ovr\Lambda_\chi}\Bigg\}~.
\eeq  
The corresponding power counting is called the  ``small scale
expansion'' (SSE). It is important to note that the mass splitting $m_\Delta - m_N$
{\em does not vanish in the chiral limit}, therefore {\em the explicit inclusion
of the delta is a phenomenological extension of CHPT}. 
More recently, it was realized that for certain considerations/processes 
a Lorentz-invariant formulation of baryon chiral perturbation
theory is advantageous, particular related to the spin sector probed e.g. in
doubly virtual Compton scattering. A particularly elegant scheme to perform covariant
calculations is the so--called ``infrared regularization'' (IR) of~\cite{Becher:1999he}.
In \cite{Bernard:2003xf} 
a consistent extension of the infrared regularization method in the presence
of  spin-3/2 was given. It was in particular shown that the contribution of the 
non-propagating spin-1/2  components of the Rarita-Schwinger field can be
completely absorbed in the polynomial terms stemming from the most general
effective chiral Lagrangian (see also the recent work
in~\cite{Hacker:2005fh}).

Here, I report on some results obtained in~\cite{Bernard:2005fy}.
In that paper,  the  nucleon and the delta mass as well as the corresponding 
sigma terms were analyzed to  fourth order in the small parameter $\ve$
(for related early work on this problem see \cite{Zanotti:2004qn}).
These explicit representations of 
$m_N$ and $m_\Delta$ serve as chiral extrapolation functions to analyze the
lattice data from the MILC collaboration \cite{Bernard:2001av} for unphysical pion
masses as low as $M_\pi \simeq 350$~MeV (note that we treat the SU(3) data as
if they were SU(2) - the same comments as in Sec.~\ref{sec:mN} apply).
Clearly, such extrapolation functions based on chiral
perturbation theory cease to make sense at too large values of the quark (pion)
masses, but as we will demonstrate later, we can nicely capture the trend of
these data. Before proceeding, it should be stressed that in the theory with
explicit spin-3/2 degrees of freedom one has more oeprators and thus more
LECs than in nucleon CHPT -- similar to the case when going from SU(2) to SU(3)
in CHPT discussed earlier.

\noindent
The corresponding expansions to ${\mathcal O}(\ve^4)$ for the nucleon and the
delta mass take the form~\cite{Bernard:2005fy}
\beqa
m=m_0-4c_1M_\pi^2-4M_\pi^2\Delta_0 D_1 - D_2\Delta_0^3-4 e_1(\lambda)  M_\pi^4 
 -E_1 \Delta_0^4 -4 M_\pi^2 \Delta_0^2 E_2  + m^{{\rm N-loop}}
+  m^{{\rm \Delta-loop}}~,\label{mndelta} \nonumber\\
m_\Delta=m_\Delta^0-4a_1M_\pi^2-4M_\pi^2\Delta_0D_{1}^\Delta
    -D_{2}^\Delta\Delta_0^3 -4 e_1^\Delta(\lambda) M_\pi^4  
-E_1^\Delta \Delta_0^4 -4 M_\pi^2 \Delta_0^2 E_2^\Delta
+m_\Delta^{{\rm N-loop}}+m_\Delta^{{\rm
    \Delta-loop}}~,\label{mdelta}\nonumber\\
&&
\eeqa
where I have made explicit the contribution from the various
contact interactions. Consider first the nucleon. Its representation
looks very similar to the one given in Eq.~(\ref{mN}). There are, however,
some important differences. First,  the appearance of the delta
in the loops also generates additional renormalizations $\sim\Delta_0^n$ ($n \ge
1$) (accompanied by combinations of dimension three ($D_{1,2}, D_{1,2}^\Delta$
and dimension four ($E_{1,2}, E_{1,2}^\Delta$) LECs) that can be absorbed in the 
LECs of the theory without delta (this is discussed in detail 
in~\cite{Bernard:1998gv}) . These LECs  can be chosen in  such away
that the delta contribution to the nucleon mass starts at ${\mathcal
  O}(M_\pi^4)$. Second, since there is an explicit delta-loop contribution,
the values of the LECs $c_i$ differ from the ones determined earlier by
the explicit delta contribution that has been worked out
in~\cite{Bernard:1996gq}. Third, the delta loops introduce new parameters,
in particular the leading axial nucleon-delta coupling called $c_A$ and the
dimension two couplings $b_1$ and $b_6$ from the effective $\pi N\Delta$
Lagrangian appear in the expression for $m_N$, utilizing
\begin{eqnarray}
{\cal L}_{\pi N\Delta}^{(1)}&=&c_A\left\{\bar{\psi}_i^\mu w_\mu^i \psi_N
                               + {\rm h.c.}  \right\}~, \label{s12321} \\
{\cal L}_{\pi N\Delta}^{(2)}&=&\bar{\psi}_i^\mu\left\{
                               b_3\,i\,w_{\mu\nu}^i\,\gamma^\nu
                               +\frac{b_6}{m_0}\,i\,w_{\mu\nu}^i\,i\,D^\nu
                               + \ldots \right\} \psi_N+ {\rm h.c.}~,
\end{eqnarray}
with $w_\mu^i = \langle\tau^i u_\mu\rangle/2$ and 
$w_{\mu\nu}^i =  \langle\tau^i \left[D_\mu,u_\nu\right]\rangle/2$,
where $\tau^i,\,i=1,2,3$ denote the Pauli matrices in isospin space.
Here, ${\psi}_i^\mu$ denotes the spin-3/2 field (for more details,
see~\cite{Bernard:1996gq}). The corresponding fourth-order diagrams 
for the delta self-energy are shown in the left panel of Fig.~\ref{fig:delta}.
In particular, one has dimension two and four LECs corresponding to the
ones in the nucleon mass case, for completeness I display the pertinent
terms of the dimension two effective Lagrangian,
\beq
{\cal L}_{\pi\Delta}^{(2)}
= -\,\bar{\psi}^\mu_i\left\{\left[a_1 \langle
\chi_+\rangle -\frac{a_2}{4 m_0^2}\left\{\langle u_\alpha u_\beta
\rangle D^\alpha D^\beta+ {\rm h.c.}\right\}
+\frac{a_3}{2} \langle u^2\rangle + \ldots\right]g_{\mu\nu}\delta^{ij}
+ \ldots \right\} \psi^\nu_j~,
\eeq
so that the $a_i$ of the delta correspond to the $c_i$ of the nucleon
and similarly $e_1^\Delta$ is a combination of dimension four LECs as
it is the case for $e_1$ discussed earlier.
 
We are now in the position to analyze the pion mass dependence of the nucleon
and delta mass given in Eqs.~(\ref{mndelta},\ref{mdelta}). They
contain a certain number of LECs, some of which are (not very accurately)
known from the study of pion-nucleon scattering in the heavy baryon SSE
\cite{Fettes:2000bb}.  In addition to the known values at the physical point 
we take the data from MILC \cite{Bernard:2001av}
for the nucleon and the delta as function of the pion mass and try to
describe these with LECs of natural size. Such a description is indeed
possible, as shown in Fig.~\ref{fig:delta}. So we do not intend least-square
fits here but rather try to find out whether the existing data shown in 
this figure can be consistently described by our mass formulas with LECs of
natural size. We stress again that a more refined analysis of
e.g. pion-nucleon scattering in the covariant SSE is mandatory to put
stringent constraints on certain combinations of the LECs.
The parameters corresponding to these curves are:
\begin{figure}[tb]
\flushleft{
\fbox{\includegraphics*[width=6.5cm]{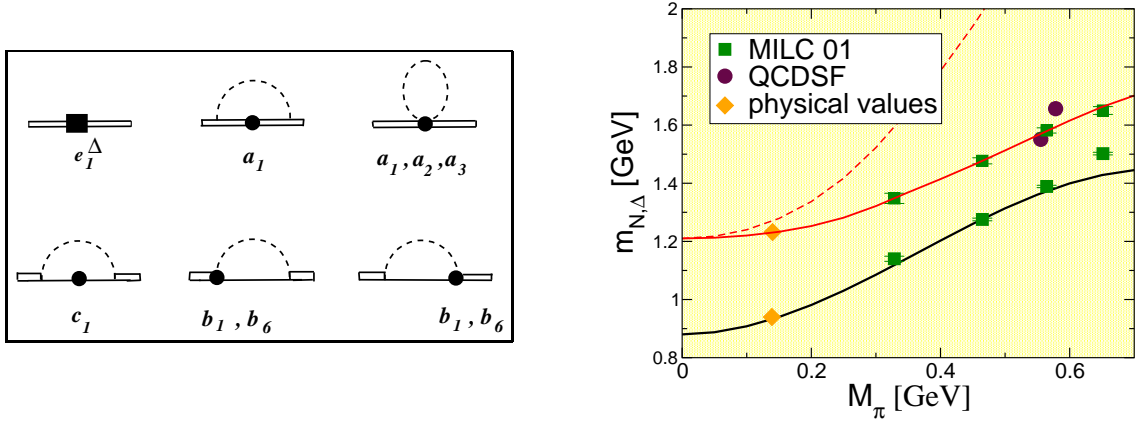}}
}

\vspace{-5.0cm}

\flushright{
\includegraphics*[width=7cm]{mdelnew.eps}
}
\caption{Left panel: Fourth order contribution to the delta mass.
Shown are the contact interactions, pion loops (dashed lines) with intermediate
nucleons (single lines) and intermediate deltas (double lines). The relevant
LECs are also shown. Right panel: The nucleon mass (black line) and the 
(real part of the) delta mass (red line)
as a function of the pion mass. The filled diamonds denote their physical
values at  the physical pion mass. The dashed line is the chiral extrapolation
for the $\Delta$ based on SU(6) as explained in the text. The filled squares
and circles are the MILC data \protect\cite{Bernard:2001av}. The filled triangles are
the recent data from QCDSF \protect\cite{Schierholz}.}
\label{fig:delta}
\end{figure}
\noindent
(i) the $N\Delta$ mass splitting in the chiral limit, 
$\Delta_0  = 0.33\,$GeV, indicating a slightly {\it larger} $N\Delta$ mass
splitting in the chiral limit than at the physical point. 
(ii) The LECs from the pion-nucleon Lagrangian 
are given by $c_1=-0.8\,$GeV$^{-1}$ (which is within the
uncertainty of the values determined in e.g. \cite{Buettiker:1999ap}) and 
$e_1 = c_2 = 0$, $c_3=0.5\,$GeV$^{-1}$. The small values of $c_{2,3}$ are
consistent with resonance saturation studies of \cite{Bernard:1996gq} and the
fits in \cite{Fettes:2000bb}. The fourth-order LEC $e_1$ induces the largest uncertainty
-- even a small value of $e_1$ leads to a sizeable contribution at larger pion masses. 
(iii) The three axial $N \Delta$ LECs are
$c_A=1.1$, $b_3 = 0.75$ GeV$^{-1}$ and $b_6 = -0.75$. (iv) The 
axial $\Delta \Delta$ coupling is found to be $h_A=2$, which is not far from the
SU(6) or large-$N_C$ value $h_A = 9 g_A/5 = 2.28$. Furthermore we get 
$e_1^\Delta =-1\,$GeV$^{-3}$, $a_1 = -0.3\,$GeV$^{-1}$, 
$A/4 +B  = a_2/4 + a_3 + \ldots =0.5\,$GeV$^{-1}$. 
These are all natural values. It is interesting to note that $a_1$ is markedly
smaller than $c_1$, although both couplings should be equal in the SU(6)
limit. Still, it is interesting to study the strict SU(6)
limit. In that case, one would have $a_1 = c_1 = -0.8\,$GeV$^{-1}$ and $h_A = 2.28$.
As can be seen from the dashed line in 
Fig.~\ref{fig:delta}, the assumption of strict SU(6) symmetry is clearly at odds
with the MILC data, indicating that $a_1$ and $c_1$ indeed seem to have
different values. Also shown in Fig.~\ref{fig:delta} are the recent QCDSF 
data for $m_\Delta$,
which were not used in the fit but are nicely consistent with our
extrapolation function. Note also that the  QCDSF data are based on two-flavor
simulations and are not very different from the MILC data in the region of
overlap. This further supports our assumption on the treatment of the MILC data.
We stress again that the resulting values of the LECs are to
be considered indicative and a more detailed 
analysis employing also constraints from other physical processes should follow.
From the small value of the LEC $a_1$ one immediately deduces that the
$\pi\Delta$ sigma term appears to be significantly {\it smaller} than 
its nucleon cousin because
at leading order in the quark mass expansion we have $\sigma_{\pi N} = 
-4c_1 M_\pi^2+...$ and $\sigma_{\pi \Delta} =-4a_1 M_\pi^2+...$.
It is clear that this interesting observation deserves further study. 
Finally we note that the sigma term for 
the nucleon resulting from this ``rough" fit is found as
\beq
\sigma_{\pi N} = 48.9~{\rm MeV}~, 
\eeq
to order $\ve^4$. We note that this  value is consistent with the 
classical result of
Ref.~\cite{Gasser:1990ce}, 
which was confirmed in \cite{Buettiker:1999ap} in a heavy baryon
CHPT analysis of pion-nucleon scattering and
in \cite{Procura:2003ig} in a CHPT analysis of lattice data.
It is also in agreement with the recent
CHPT analysis of the three--flavor MILC data, see~\ref{sec:mN}.  For the
$\pi\Delta$ sigma term we get $\sigma_{\pi\Delta} = 20.6\,$MeV. 
Note that model studies give slightly larger values between 28 and 35\,Mev
\cite{Lyubovitskij:2000sf,Cavalcante:2005mb}.
Again,
these results need to be refined and bolstered by more detailed precise
fits to the lattice data including also error and correlation analysis
including also lattice data on other observables - the mass data alone are
not sufficient to precisely pin down all parameters. Such an analysis,
however, is not yet available. Note also that chiral extrapolation functions for the
baryon octet and decuplet masses in partially quenched QCD based on an EFT
with deltas are discussed in~\cite{Walker-Loud:2004hf,Tiburzi:2004rh}.

\section{Quark mass dependence of nuclear forces}
\label{sec:NN}
Because of the smallness of the up and down quark masses, one does not expect
significant changes in systems of pions or pions and one nucleon when the
quark masses are set to zero (with the exception of well understood chiral
singularities like e.g. in the pion radius or the nucleon
polarizabilities). The situation is more complicated for systems of two (or
more) nucleons. Here, I report on
some work \cite{Epelbaum:2002gb} that is mostly concerned with the properties of the
deuteron and the S-wave scattering lengths as a function of the quark (pion)
mass. These questions are not only of academic interest, but also of practical
use for interpolating results from lattice gauge theory. E.g. the S-wave
scattering lengths have been calculated on the lattice using the quenched 
approximation \cite{Fukugita:1994ve}. Another interesting application is related to
imposing bounds on the time-dependence of some fundamental coupling constants
from the two--nucleon (NN) sector, as discussed in \cite{Beane:2002vs}. 

\noindent
To address these issues, consider the chiral two-nucleon potential at
next-to-leading order (NLO). It is given in terms of one- and two-pion
exchanges (OPE and TPE, respectively) and four-nucleon contact interactions 
\beq
V = V_{\rm  OPE} + V_{\rm TPE} + V_{\rm contact}~.
\eeq
In this approach, we have to deal with  {\em explicit} and  
{\em implicit} quark mass dependences. This can be most easily understood
by looking at the OPE and contact interactions:
\begin{eqnarray}\label{VNN}
V_{\rm OPE} &=& \left( -\frac{g_A^2}{4F_\pi^2} + \ldots \bar{d}_{16} + \ldots \right)
\, {\vec \tau}_1 \cdot {\vec \tau}_2 \, \frac{(\vec\sigma_1 \cdot \vec q )
(\vec\sigma_2 \cdot  \vec q )}{{\vec q\,}^2 + {M_\pi^2}}\\
 V_{\rm contact} &=& C_S + M_\pi^2 \,  {D_S} + (C_T + M_\pi^2 \, 
 {D_T}) \, (\vec\sigma_1 \cdot 
\vec\sigma_2 ) + \ldots
\end{eqnarray}
The OPE exhibits both types of quark mass dependences.
The pion propagator becomes Coulomb-like in the chiral limit,
$1/({\vec q\,}^2 + M_\pi^2) \to 1/ {\vec q\,}^2$.
The implicit pion mass dependence enters at NLO through the pion--nucleon
coupling constant (note that the quark mass dependence of the nucleon
mass only enters at NNLO) expressed through the pion mass dependence of $g_A/F_\pi$
in terms of the quantity
\beq
\label{deltaCL}
\Delta = \left( \frac{g_A^2 }{16 \pi^2 F_\pi^2} - \frac{4 }{g_A}
\bar{d}_{16} + \frac{1}{16 \pi^2 F_\pi^2} \bar{l}_4 \right) 
(M_\pi^2 - \tilde M_\pi^2) - 
\frac{g_A^2 \tilde M_\pi^2}{4 \pi^2 F_\pi^2} \ln \frac{\tilde M_\pi}{M_\pi} \, .
\eeq
Here $\bar l_4$ and $\bar d_{16}$ are LECs related to 
pion and pion--nucleon interactions, and the value of the 
varying pion mass is denoted
by $\tilde M_\pi$ in order to distinguish it from the physical one denoted by
$M_\pi$. In particular, $\bar d_{16}$ has been determined in various fits to
describe $\pi N \to \pi\pi N$ data, see \cite{Fettes:1999wp}.
\begin{figure}[tb]
\centerline{\psfig{file=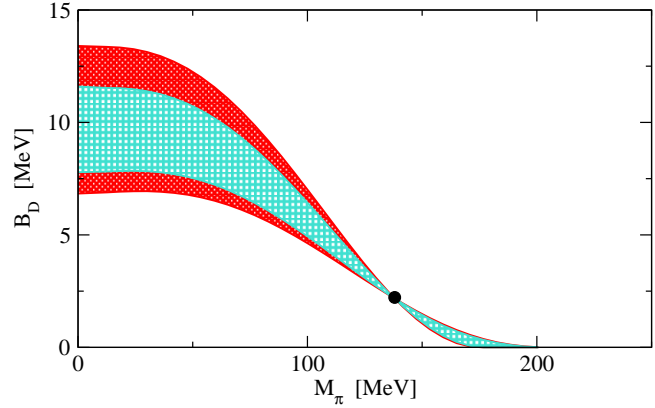,width=9.5cm}}
\caption{Deuteron BE versus the pion mass. The shaded areas show the 
allowed values. The light shaded band corresponds to our main result with  
the uncertainty due to the unknown LECs $D_{S,T}$.
The dark shaded band gives the additional uncertainty due to the 
uncertainty of $\bar d_{16}$.
The heavy dot shows the BE for the physical case $\tilde M_\pi = M_\pi$.}
\label{figBE}
\end{figure}
%
\noindent
The four-nucleon contact terms exhibit an explicit quark mass dependence,
These NLO $M_\pi^2$ corrections to the leading operators $\sim C_S, C_T$
are parameterized by the LECs $D_{S}$ and $D_T$, respectively. 
These LECs can at present only be estimated using dimensional analysis 
and resonance saturation~\cite{Epelbaum:2001fm}. 
For the explicit expressions of the TPE, see~\cite{Epelbaum:2002gb}.

\begin{figure}[htb]
\begin{center}
\epsfig{file=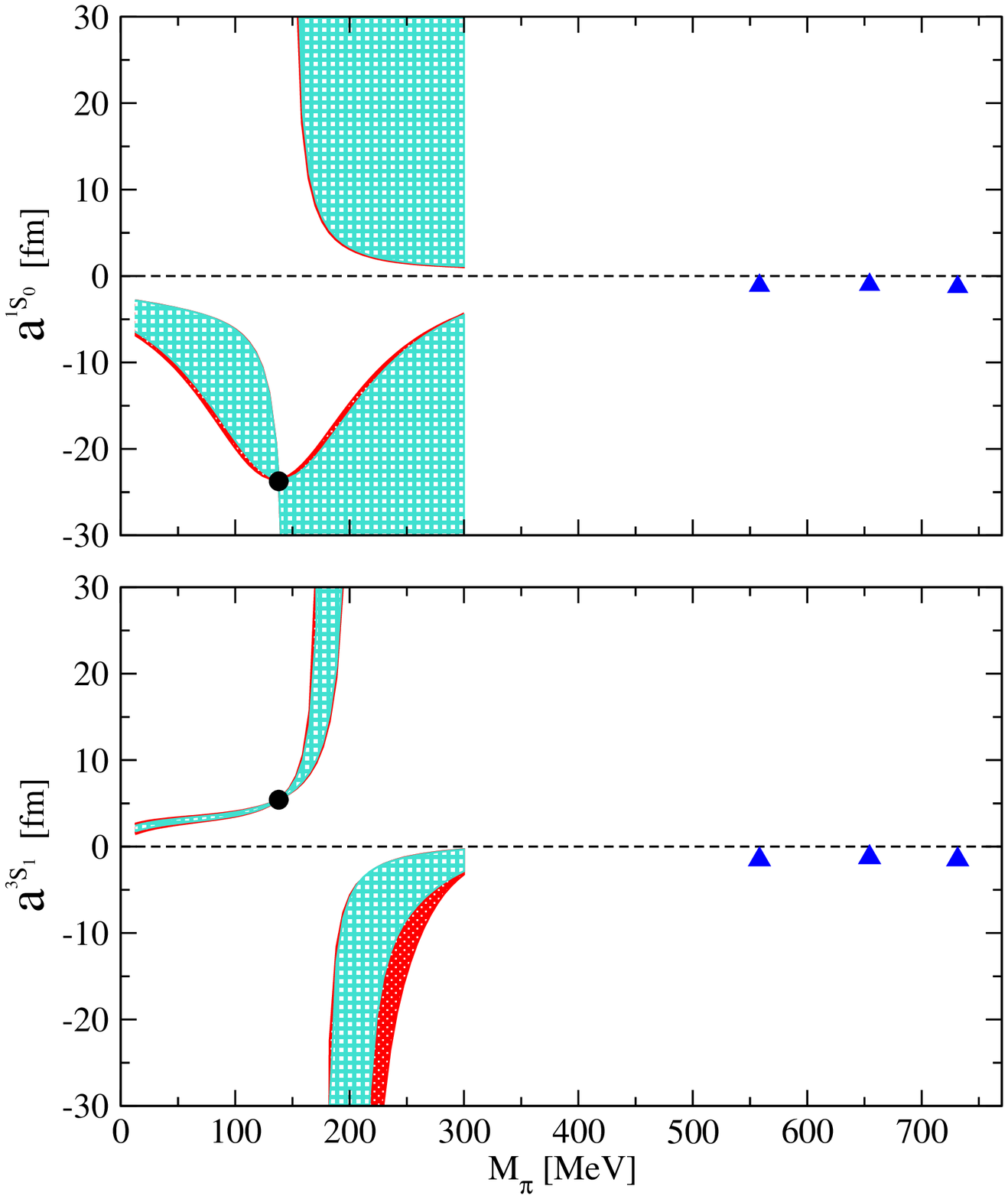,width=5.9cm} \hspace{1cm}
\epsfig{file=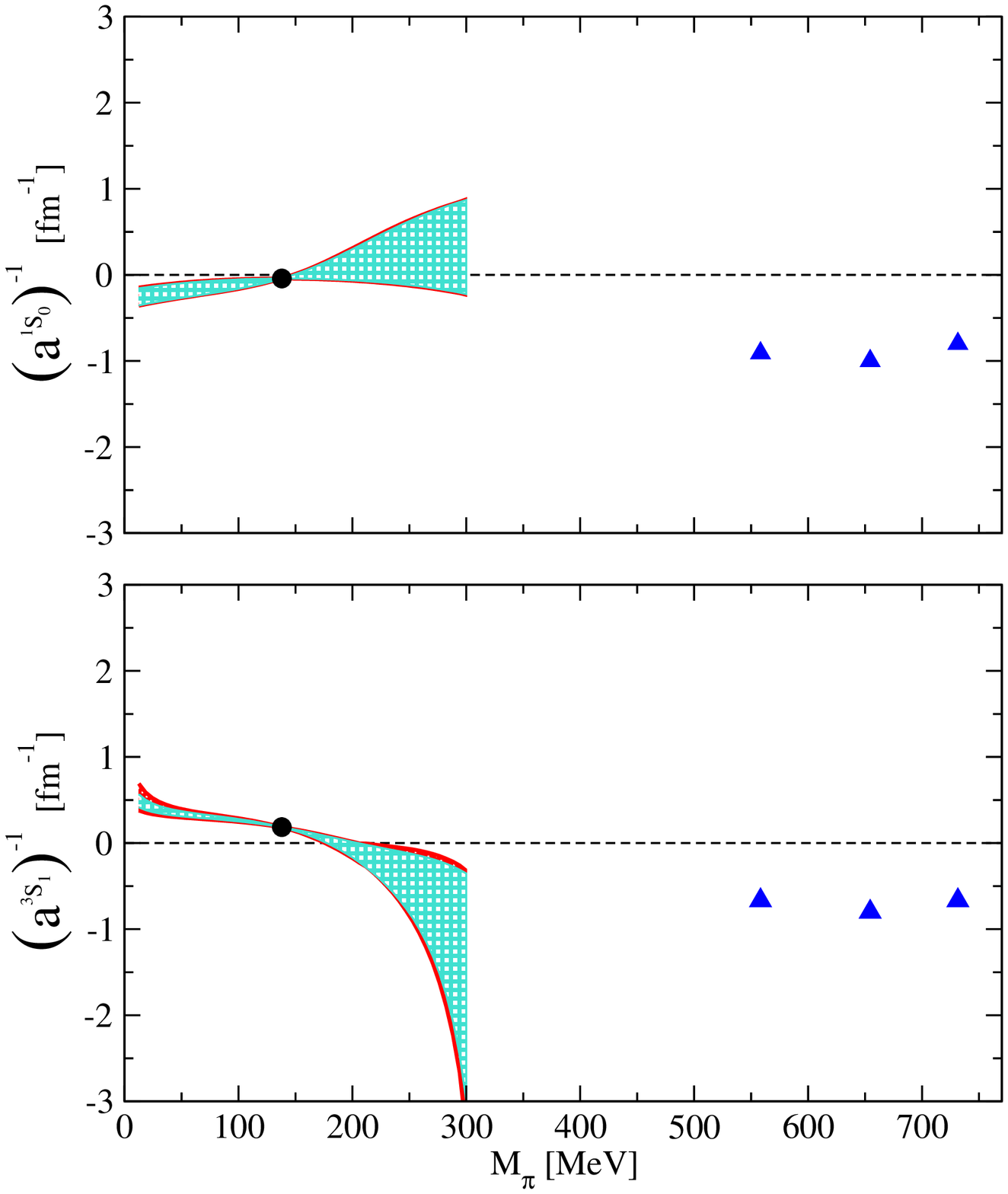,width=5.9cm}
\caption{Left panel: The S--wave scattering lengths as functions 
of pion mass.
The shaded areas represent the allowed values according to our analysis.
The light shaded band corresponds to our main result with  
$\bar d_{16} = -1.23$ GeV$^{-2}$ and the uncertainty due to the 
unknown LECs $\bar D_{S,T}$.
The dark shaded band gives the uncertainty if, in addition to variation of 
$\bar D_{S,T}$, 
the LEC $\bar d_{16}$ is varied in the range from 
$\bar d_{16} = -0.91$ GeV$^{-2}$ to $\bar d_{16} = -1.76$ GeV$^{-2}$.
The heavy dots corresponds to the values in the real world. The triangles 
refer to lattice QCD results from \protect\cite{Fukugita:1994ve}.
Left panel: The S--wave inverse scattering lengths as functions of 
the pion mass. Notation as before.}
\label{figaS}
\end{center}
\end{figure}

\noindent
It is now straightforward to solve the Lipp{\-}mann-Schwinger equation for the
bound and scattering states utilizing the properly regularized chiral
EFT potential, Eq.~(\ref{VNN}).
The deuteron binding energy (BE) as a function of the pion mass 
is shown in Fig.\ref{figBE}.
We find that the deuteron is stronger bound in the chiral limit than in the
real world,
\beq
B_{\rm D}^{\rm c.l.} =  9.6 \pm 1.9 {{+ 1.8} \atop  {-1.0}}~{\rm  MeV}~,
\eeq
where the   first indicated error refers to the uncertainty in the value 
of $\bar D_{^3S_1}$ and $\bar d_{16}$ being set to its average value 
while the second indicated error shows the additional uncertainty due 
to the uncertainty in the determination of $\bar d_{16}$.
Note that in ~\cite{Beane:2002xf} a larger variation of the LECs $D_{S,T}$
was considered, leading to the possibility that the deuteron can become
unbound in the chiral limit. Also, for massless pions, the higher partial
waves are no longer suppressed by the centrifugal barrier, $\delta_l (k)
\sim k$ ($l = 1,2,\ldots$) due to the Coulomb-like pion propagator.
Consequently, one could also have bound
states in  partial waves other than $^3S_1-^3D_1$ (the deuteron channel).
However, we did not find other bound states. Last but not least,
we found smaller (in magnitude) and more natural values for the two 
S--wave scattering lengths in the chiral limit,
\beq
a_{\rm c.l.} (^1S_0) = -4.1 \pm 1.6 {{+ 0.0} \atop 
{-0.4}} \,{\rm fm} \, ,\quad  
a_{\rm c.l.} (^3S_1) = 1.5 \pm 0.4 {{+ 0.2} \atop 
{ -0.3}}\, {\rm fm}\, .
\eeq
The pion mass dependence of the scattering lengths and their inverse is
shown in Fig.~\ref{figaS}. As can be seen from that figure, 
one needs lattice data for pion masses below
300 MeV to perform a stable interpolation to the physical value of $M_\pi$.
We conclude that nuclear physics in the chiral limit is much more natural than 
in the real world. 

\section{Summary and outlook}

In this talk, I have discussed the foundations and some applications
of the quark mass dependence of baryon properties, based on chiral
perturbation theory and extensions thereof. This can be summarized as
follows:

\begin{itemize}
\item Baryon CHPT is a mature field in the light quark sector. For the
light quark flavors up and down, the explicit symmetry breaking is 
generally small
at the physical point. This leads to the observation that fourth-order,
one-loop calculations provide accurate and unambiguous extrapolation
functions for lattice QCD. In the three flavor sector, the situation
is less satisfactory, nonetheless a variety of observables can be studied
without resorting to resummation techniques, coupled channel analysis or
models. The formalism for a covariant inclusion of 
matter fields for spin-1/2 and spin-3/2 fields exist. Because of decoupling,
only the
ground state baryons can contribute to the leading non-analytic terms
in the chiral expansion. If one explicitely includes the spin-3/2 decuplet,
one must therefore respect decoupling, as discussed in Sec.~\ref{sec:cl}.

\item It is important to realize that observables are linked by general 
operator structures. Stated differently, the low-energy constants (LECs) 
are universal, they contribute to different processes. For that reason
and given the present scarcity of low pion mass lattice simulations, one
should perform global fits to observables combined with input from
phenomenology. The quark mass dependence of any observable is given in terms
of loop contributions and local operators parameterized by class~II LECs,
the so-called symmetry breakers. The dynamical (class~I) LECs  corresponding to
operators that do not vanish in the chiral limit can best be determined
by phenomenological analysis. I have tried to stress this issue by considering
the determination of some of the dimension two pion-nucleon LECs, their
appearance in the nucleon mass and in loop corrections to the nucleon
isovector magnetic moment.  When considering three flavor CHPT, it is
important to realize that the SU(3) LECs are related to the SU(2) ones 
by matching conditions -- this allows to get some constraints when analyzing
baryon octet observables. 

\item Given the existing lattice data and our present knowledge of the
LECs and the quark mass dependence of important input parameters like
the pion decay constant or the axial-vector coupling constant, I conclude 
that CHPT gives chiral extrapolation functions with a small and moderate
theoretical error for $M_\pi \lesssim 400$ and $500$~MeV, respectively. 
However, in certain cases a smaller window of stability arises
at a given order in the chiral expansion, see e.g. the discussion
of $g_A$ in Sec.~\ref{Nprops}. One of the strengths of the effective field theory
approach is the possibility to systematically work out theoretical
uncertainties -- almost all figures shown here underline this important issue. 

\item Chiral nuclear effective field theory allows to analyze the quark mass
dependence of few-nucleon systems, some results for the two-nucleon sector
were discussed in Sec.~\ref{sec:NN}. The two-nucleon force appears to be
more natural in the chiral limit than for physical quark masses. Also, the
observation that both S-wave scattering lengths vanish simultaneously for 
$M_\pi\simeq 175\,$MeV has led to the speculation of an QCD IR limit cycle 
in the three-nucleon system~\cite{Braaten:2003eu}. Its relevance for lattice
QCD was already stressed by Ken Wilson at Lattice~2004~\cite{Wilson:2004de}. 

\item All this can be summarized in one simple sentence: To connect lattice
QCD with real QCD (nature) in well-defined and precise manner, {\em we need
lattice data at low pion masses} -- most studies performed so far (including
the ones presented here) are merely of pioneering nature, eventually paving
the way for truly {\it ab initio} calculations of baryon properties.
\end{itemize}

\section*{Acknowledgements}
I would like to thank my collaborators, first and most, V\'eronique Bernard, Thomas
Hemmert, Matthias Frink and Evgeny Epelbaum. I would also like to take the
opportunity to thank the organizers for setting up such an interesting conference.
Useful communications from Claude Bernard, Meinulf G\"ockeler and Gerrit
Schierholz are also acknowledged.

\vfill \eject

\bibliographystyle{JHEP}
\bibliography{skeleton}

\end{document}